# Electric field-controlled rippling of graphene

Zoltán Osváth,*[a,b] François Lefloch,[a] Vincent Bouchiat,[c] and Claude Chapelier*[a]

[a] SPSMS, UMR-E 9001, CEA-INAC/UJF-Grenoble 1, 17 rue des martyrs, 38054 Grenoble cedex 9, France; E-mail: claude.chapelier@cea.fr

[b] Institute of Technical Physics and Materials Science, MFA, Research Centre for Natural Sciences, Hungarian Academy of Sciences, P.O. Box 49, 1525 Budapest, Hungary; E-mail: osvath.zoltan@ttk.mta.hu

[c] Institut Néel, CNRS and Université Joseph Fourier, BP 166, 38042 Grenoble cedex 9, France

## Abstract

Metal/graphene interfaces generated by electrode deposition induce barriers or potential modulations influencing the electronic transport properties of graphene based devices. However, their impact on the local mechanical properties of graphene is much less studied. Here we show that graphene near a metallic interface can exhibit a set of ripples self-organized in domains whose topographic roughness is controlled by the tip bias of a scanning tunneling microscope. The reconstruction from topographic images of graphene bending energy maps sheds light on the local electro-mechanical response of graphene under STM imaging and unveils the role of the stress induced by the vicinity of graphene/metal interface in the formation and the manipulation of these ripples. Since microscopic rippling is one of the important factors that limit charge carrier mobility in graphene, the control of rippling with a gate voltage may have important consequences in the conductance of graphene devices where transverse electric fields are created by contactless suspended gate electrodes. This opens also the possibility to dynamically control the local morphology of graphene nanomembranes.

## 1 Introduction

The possibility to design high-speed electronic devices based on graphene has attracted intense research in recent years, due to the outstanding electronic properties of this single atomic layer thick carbon sheet.[1] Electrons and holes in graphene behave as massless Dirac fermions[2,3] and can have unprecedented high mobilities up to room temperature.[4,5] However, the carrier mobility is significantly lower in substrate-supported (SiO$_2$) graphene devices,[6,7] due to various substrate-induced effects like scattering from remote surface phonons,[7] charged impurities[8] and microscopic corrugations or ripples.[9,10]

It has been shown by scanning probe microscopy that graphene approximately follows the corrugation of the SiO$_2$ substrate[11] and the generated corrugation induces modulations in the local tunneling conductance.[12] However, recent scanning tunneling microscopy (STM) investigations of SiO$_2$-supported graphene suggest that the graphene is partly suspended, as it exhibits a corrugation period which is not induced by the substrate.[13] This has been further demonstrated by the induced motion of ripples of graphene on silicon dioxide with the help of an STM tip,[14] an electromechanical effect which has also been addressed theoretically.[15,16,17] Whereas these ripples are intrinsic to the free-standing structure,[18,19] boundary confinement can also induce large ripples in graphene, like in any strained elastic membrane.[20] In this work we investigate SiO$_2$-supported graphene in the proximity of a metallic contact, and show experimentally that rippling on the suspended areas of this strained region of graphene can be drastically enhanced and controlled by the transverse electric field induced by the voltage-biased STM tip.

**2 Experimental section**

The graphene sample was prepared by micromechanical cleavage of Kish graphite on a highly doped silicon wafer with 285 nm SiO$_2$ which was pre-marked for lithographic positioning. In order to prepare the sample for STM measurements, a first step of e-beam lithography was employed using PMMA (4%). An area of 500×500 μm$^2$ around the graphene flake (including the edges of the flake) was exposed to an e-beam dose of 250 μC/cm$^2$. After development in a solution of Methyl Isobutyl Ketone and Isopropyl Alcohol (1:3), a thin film of Nb/Au (50 nm/15 nm) metals was deposited to contact the edges of the graphene and to metalize the surface exposed by the e-beam. As a second lithographic step, Ti/Au (5 nm/25 nm) stripes of different shapes and sizes were deposited on a large (500×500 μm$^2$) area around the sample, to facilitate the location of the graphene flake with the STM tip. The thorough removal of the resist was ensured by a long lift-off time of several hours in acetone bath at 55 $^o$C. Topographic and spectroscopic measurements were performed under ambient conditions, using a home-built STM with Omicron control system.

## 3 Results and discussion

An optical image of the investigated sample is shown in Figure 1a, where the dark central region reveals the graphene, surrounded by a Nb/Au thin film (yellow area). The optical contrast reveals a darker part of the flake, which corresponds to a few-layer graphene region.

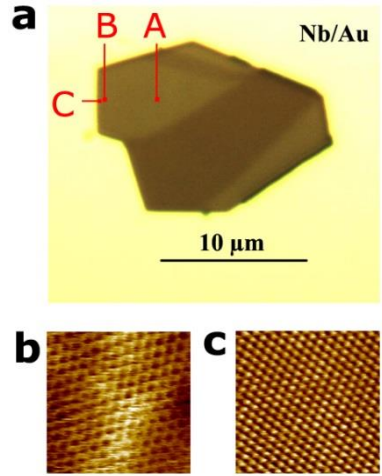

**Fig. 1** Monolayer and few-layer parts of the investigated graphene sample. (a) Optical microscope image of the graphene flake, contacted all around by Nb/Au contact. The darker areas correspond to few layers graphene whereas the lighter ones are made of a single layer of graphene. The dots labelled A, B, and C denote STM scanned areas discussed in the text. (b) and (c) atomic resolution STM images (3.5x3.5 nm$^2$, $V_b$ = 0.1 V, I = 1.5 nA) acquired on the single-layer graphene area of the sample (b) and on the few-layers graphene area (c).

High resolution STM images acquired there display the triangular lattice characteristic of graphite surfaces with Bernal (ABAB) stacking of layers (Figure 1c), whereas the hexagonal lattice of graphene monolayer was obtained outside this area (Figure 1b). The area enclosed by the metallic contact follows the shape of the flake with an overlap width of about 500 nm. The maximal length of the contacted flake (horizontal dimension, *X*) is 17 μm, while the maximal width (vertical dimension, *Y*) is 12.5 μm. We have studied three different spots on the graphene monolayer, labeled *A*, *B*, and *C* in Figure 1a, located at different distances from the metallic contact. Area *A* is 4 μm away from the metallic film, whereas areas *B* and *C* are much closer to it, at distances from the metal of 250 nm and 30 nm respectively. Details of the corrugation of graphene in areas *A* and *B* are shown in Figure 2.

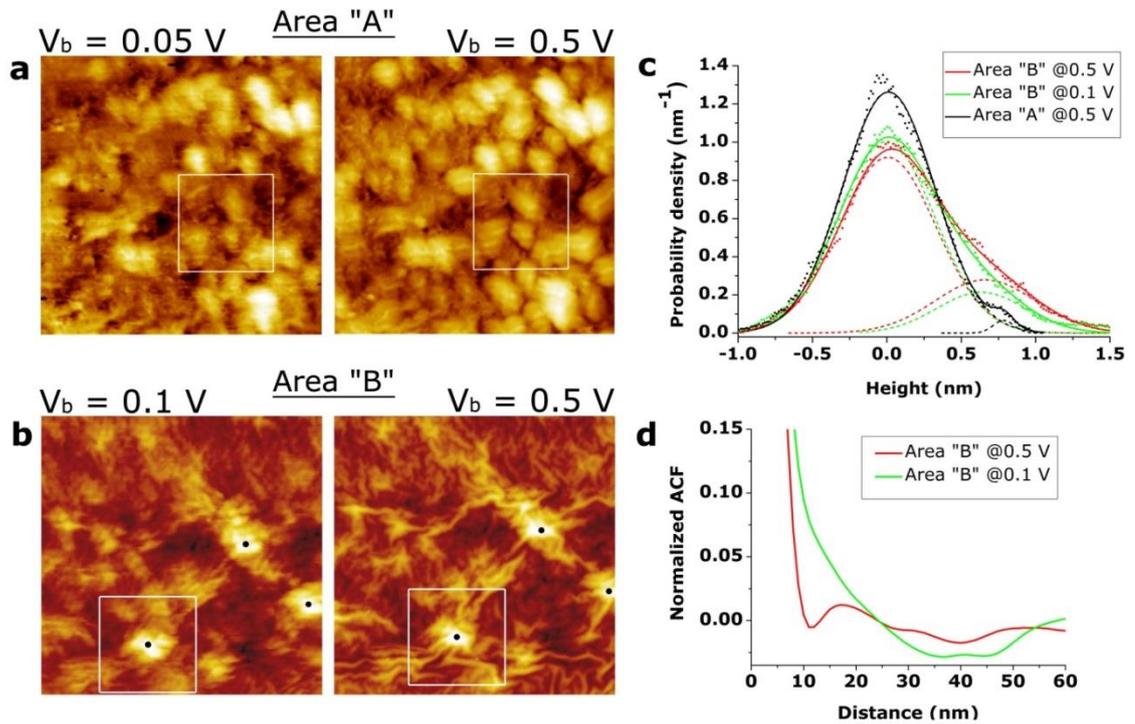

**Fig. 2** Topographic STM images (390x390 nm$^2$, I = 0.5 nA) acquired on two different areas, far and close respectively to the metal electrode. The white squares are the area where the bending energy maps of Fig. 5 are computed. (a) Images acquired in area A, far from the metallic contact (left edge of images at ~4 μm from the metal-graphene interface). (b) Images acquired in area B, close to the contact (left edge of images at ~250 nm from the interface). The black dots denote highly strained graphene regions. (c) Height histograms acquired in areas A and B. The dots are the experimental data. They are fitted by the sum (solid lines) of two Gaussian distributions (dashed lines). (d) Radial autocorrelation function (ACF) of the filtered images (1st order Butterworth high-pass Fourier filter in order to remove the large wavelength corrugation induced by the underlying SiO$_2$ substrate).

Images of Figure 2a have been acquired with two different bias voltages $V_b$, on area *A*. Both images look similar and display an isotropic large wavelength roughness mainly induced by the SiO$_2$ substrate.[13] A much more pronounced set of ripples is visible in the images of Figure 2b which have been acquired on area *B*. These ripples are superimposed to the large wavelength roughness (see supplementary information). Ripples also develop in the shape of a star centered on hills (pointed out by black dots). These localized structures are reminiscent of stress focusing which spontaneously appears during the crumpling of thin sheets.[21] Strikingly these structures become much sharper for higher voltage biases, the apparent corrugation being locally enhanced by 0.2 nm. A similar, but back-gate voltage dependent deforming effect was recently observed in suspended graphene sheets.[22] These observations are more quantitatively underlined in Figure 2c, where height histograms are shown. Each histogram can be described by the sum of two Gaussian distributions. The main contribution corresponds to the corrugation induced by the silica substrate and has similar width for both images. The second Gaussian distribution is due to the ripples, which develop at the highest and suspended parts of the graphene. Whereas on area *A* the amplitude of this second Gaussian is only 0.06 and is weakly dependent on the bias voltage (not shown), it reaches 0.21 and 0.27 at $V_b$ = 0.1 V and 0.5 V respectively, for the images acquired on area *B*. This shows that the strain generated by the metal deposited on top of the graphene induces some buckling nearby the graphene/metal interface. These buckled ripples are suspended areas which can be easily pulled by the STM electric field. A field-induced correlation distance $R_c$ between 15 and 25 nm can be computed from the normalized autocorrelation functions (ACF) (Figure 2d).

Next, we investigated the area *C*, the closest spot to the metallic contact. Figure 3a shows topographic STM images performed with different bias voltages.

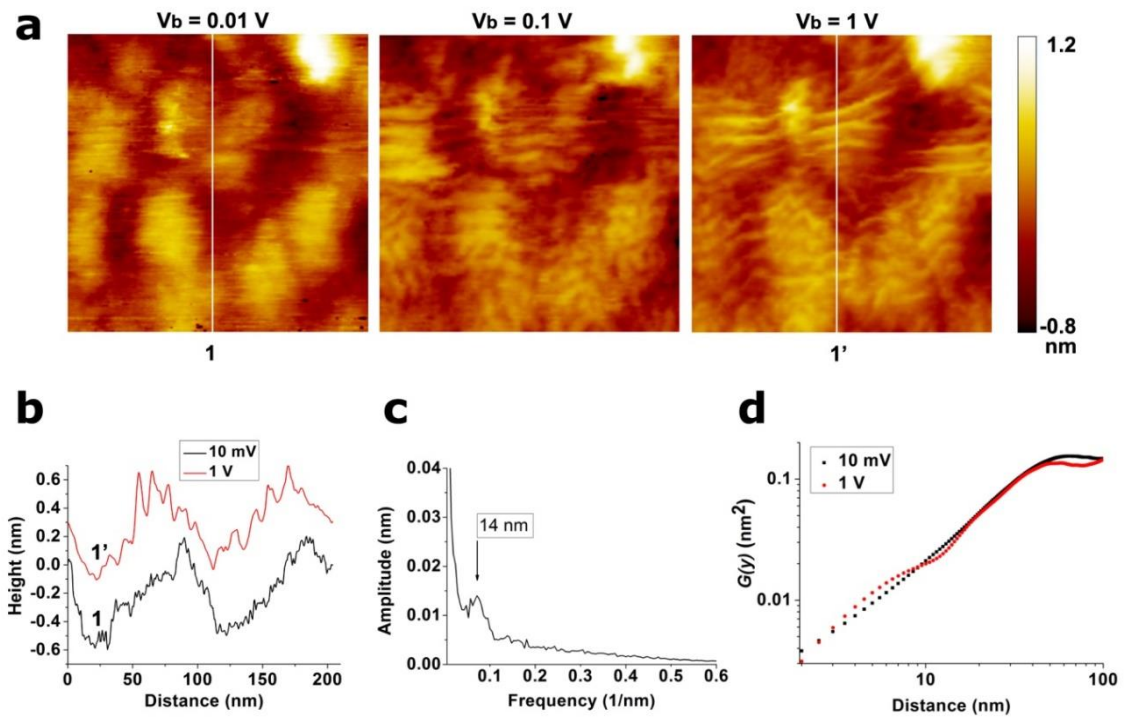

**Fig. 3** (a) Topographic STM images (210x210 nm2, I = 0.5 nA) performed in area C using different bias voltages. The left border of images is about 30 nm away from the graphene-metal interface. (b) Height profiles taken along the white lines drawn on figures (a). (c) FFT along the Y direction of the difference of the images acquired at 1V and 10 mV. (d) Height height correlation functions of the image acquired at 10 mV and 1V.

Here, in contrast with area B, there is no crumpling of the graphene membrane and no conical dislocations are visible. For $V_b$ = 10 mV, the large wavelength corrugation induced by the substrate is dominant. Nevertheless, at 100 mV a shorter period rippling along the *Y* direction is observed which is further enhanced as $V_b$ increases to 1 V. At this voltage, the ripple amplitude reaches 0.3 ± 0.05 nm (see height profiles of figure 3b). Such a pattern corresponds to parallel wrinkles, orthogonal to the graphene-metal interface, and have already been observed in various kinds of stressed membranes and in graphene as well.[20,23] This anisotropy is due to the metallic contact which imparts a shear to the graphene membrane and facilitates a rippling direction which is parallel to the metallic edge.[23,24] If one subtracts the images taken at 1V and 10 mV and performs the Fast Fourier Transform of this difference along the rippling direction, one obtains a rippling period $R_p$ of 14 nm (see Figure 3c), close to $R_c$. Finally, the height-height correlation functions $G(y) = \langle(h(y_0 + y) - h(y_0))^2\rangle$ are shown in Figure 3d. For $V_b$ = 10 mV, $G(y)$ increases as $y^{2H}$, where *2H* = 1.10 ± 0.016 characterizes the fractal dimension of the roughness when the graphene flake roughly follows the corrugation of $SiO_2$.[11] In contrast, at $V_b$ = 1 V the correlation function is no longer linear in a log-log representation, because of the induced ripples. For small distances (< 5 nm) the slope of Log ($G(y)$) can be approximately fitted with *2H* ≅ 1.6. $G(y)$ saturates for $y \approx$ 10 nm (< $R_p$) and merges with the height-height correlation function measured at low bias when *y* gets larger than 20 nm (> $R_p$). This demonstrates that for distances shorter than $R_p$, the fractal dimension of rippling gets closer to *2H* = 2, as expected for free flexible membranes.

The observed rippling enhancement is mainly a true topographic effect, although we cannot rule out a small density of states contribution. In STM imaging, the apparent corrugation of a surface is the convolution of topographic roughness and spatial fluctuations of the electronic density of states (DOS) according to the relationship:

$$I \propto e^{-\frac{2}{\hbar}\sqrt{2m\phi}d} \int_{E_F}^{E_F+eV} \rho(E + eV)\, dE, \quad (1)$$

where *I* is the tunnel current, $\phi$ the average work function of sample and tip, *V* the bias voltage, $E_F$ the Fermi energy, *d* the tip-sample distance and $\rho$ the local DOS of the sample. Thanks to its exponential contribution to the current, *d* is often the most important parameter and generally the only considered one. However, for atomically flat surfaces, a voltage sensitivity corrugation is generally attributed to $\rho(E)$. For example it is known that in graphite or in graphene, a voltage dependent moiré pattern can be observed with an STM.[25] Therefore, in the voltage dependent images reported in this work, the only way to discriminate unambiguously a true height variation from a DOS effect is to measure *dI/dV(V)* spectra. Indeed, according to (1), *dI/dV(V)* and $\rho(eV)$ are proportional quantities. We present in Figure 4b spectra acquired along a line crossing two ripples (see STM image in Figure 4a).

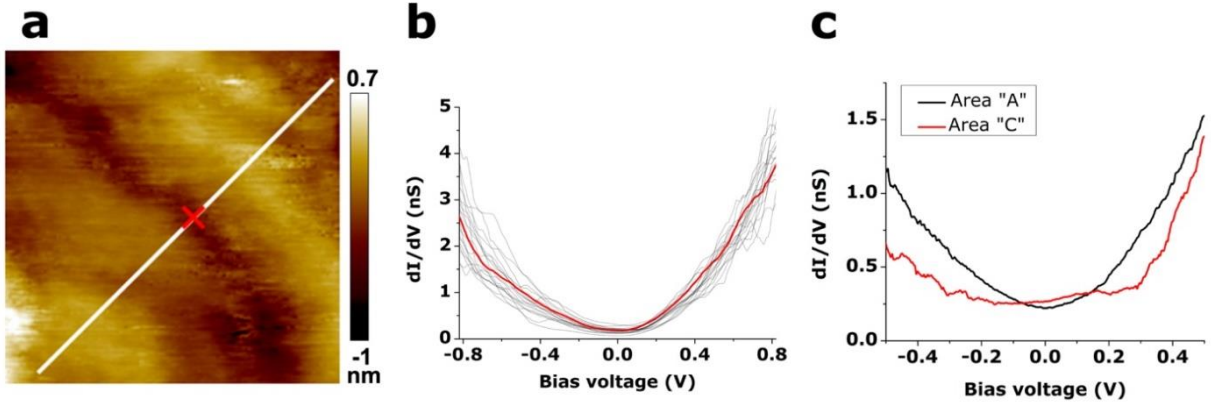

**Fig. 4** Spectroscopy measurements acquired on ripples and on strain-free regions. The spectra displayed in b) have been taken along the white line drawn on the STM image a) (80×80 nm$^2$). There is no significant difference between the spectra acquired on the two ripples or in between. The red cross in a) corresponds to the position where the red spectrum displayed in figure b) was measured. c) Spatially averaged spectra acquired on a strain-free graphene, far from the metallic contact (area A, black curve) and on a strained graphene, close to the metallic contact (area C, red curve).

We do not see any significant difference for spectra acquired on top of a ripple and in between, where the graphene sticks to the substrate. There is therefore no spectral feature at any specific voltage which could explain the observed 0.2 nm modulation of the apparent corrugation in the images acquired at 1V. Indeed, if such a variation was not a true topographic modulation, it would require a swing in the DOS amplitude of more than one order of magnitude, which is not what we observe.

Moreover, we acquired 25×25 spectra from -1 V to 1 V on a grid of 80×80 nm$^2$ on a strain-free graphene, far from the metallic contact (area *A*) and on strained graphene, close to the metallic contact (area *C*). The averaged curves are presented in Figure 4 c. It is important to note that in these spectra the fluctuations induced by substrate surface corrugations[12] and electron-hole puddles[26,27] are averaged out. One can observe that the shape of the *dI/dV(V)* curve obtained close to the metallic contact is different from the typical V-like spectrum measured far from the interface. It is a U-shaped spectrum which is characteristic to strained regions,[12] and is consistent with the observed presence of ripples. The spectrum close to the metal interface is more asymmetric, with higher local density of states above the Fermi energy, indicating *n*-type doping. This is consistent with the fact that niobium has a lower work function (4.37 eV)[28] than graphene (~4.5 eV).[29,30] This doping from a metal exists only in a region close to the metal-graphene interface. Calculations show however that this region can actually be quite large, contact induced states can penetrate a distance of ~$W/2$ into the graphene sheet,[31] where $W$ is the width of the metal contact. Photocurrent measurements[32] also demonstrate that the influence of a metal on graphene can extend to more than 450 nm (known as metal-controlled graphene region). Consequently, we assume that more charge is present close to the contact and this can have a contribution in the observed ripple enhancement due to the increased electric force.

At this stage it is interesting to compare the different energy scales which come into play. The electrostatic energy per unit area can be roughly estimated using the plate capacitor model:

$$E_{el} = \frac{1}{2} \frac{\epsilon}{d} V_b^2 \ ,$$

where $\epsilon$ is the vacuum permittivity, $d = 1$ nm is the typical tip-sample distance and $V_b$ the voltage bias. This gives an electrostatic energy of a few mJ/m$^2$ at 1V. This value is far less than the adhesion energy of graphene on SiO$_2$ of 450 mJ/m$^2$.[33] As a consequence the electric field cannot de-bond a strain-free graphene-SiO$_2$ interface. The two other relevant energies are the out-of-plane bending energy and the in-plane stretching energy of graphene.[34] However, due to the extremely high in-plane elastic modulus C = 340 N/m, we only consider the flexural deformation in our analysis. We estimate the bending energy with the formula:[34]

$$E_b = \iint \frac{D}{2} \left[ \left( \frac{\partial^2 z}{\partial x^2} + \frac{\partial^2 z}{\partial y^2} \right)^2 + 2(1-v) \left( \left( \frac{\partial^2 z}{\partial x \partial y} \right)^2 - \frac{\partial^2 z}{\partial x^2} \frac{\partial^2 z}{\partial y^2} \right) \right] dxdy, \quad (2)$$

where $z(x,y)$ is the local corrugation, $D = 2.25 \times 10^{-19}$ $Nm$ is the bending rigidity[35] and $v = 0.165$ is the Poisson's ratio of graphene.[36,37] In Figure 5, we mapped the integrand of (2) in the areas delimited by the white squares drawn in Figure 1. In area *A*, (Fig. 5a), one can observe that at high voltage bias the bending energy is concentrated along well defined stripes. These stripes correlate well with topographic features which firmly bound the graphene to the SiO$_2$ substrate. In the rest of area *A*, the bending energy is lowered by the electric field. This apparent smoothing can be explained by the dynamic lifting of the graphene as described in Ref. 14. In this mechanism the lifted part of the graphene moves simultaneously with the STM tip resulting in a reduced curvature of the image. This results in an apparent lowering of the averaged bending energy $E_b/S$, where $S$ is the total area. We found $E_b/S$ equals to 0.13 mJ/m$^2$ and to 0.09 mJ/m$^2$ at 0.05 V and 0.5 V, respectively. Note that this dynamic lifting is prohibited at the white stripes positions where the graphene keeps sticking to the substrate. A different physical picture arises in area *B*. As shown in Fig. 5b, the pre-existing ripples prevent the sliding of the crumpled graphene sheet and the electric field can only locally pull the stressed surface around the hillocks pointed out by the dark dots in Fig. 1b, and Fig 5b. In this case, the apparent averaged bending energy is increased from 0.13 mJ/m$^2$ at 0.1 V to 0.18 mJ/m$^2$ at 0.5 V. It is worth noticing that the field-enhancement of the rippling is fully reversible and does not lead to permanent and static deformation of the graphene sheet.

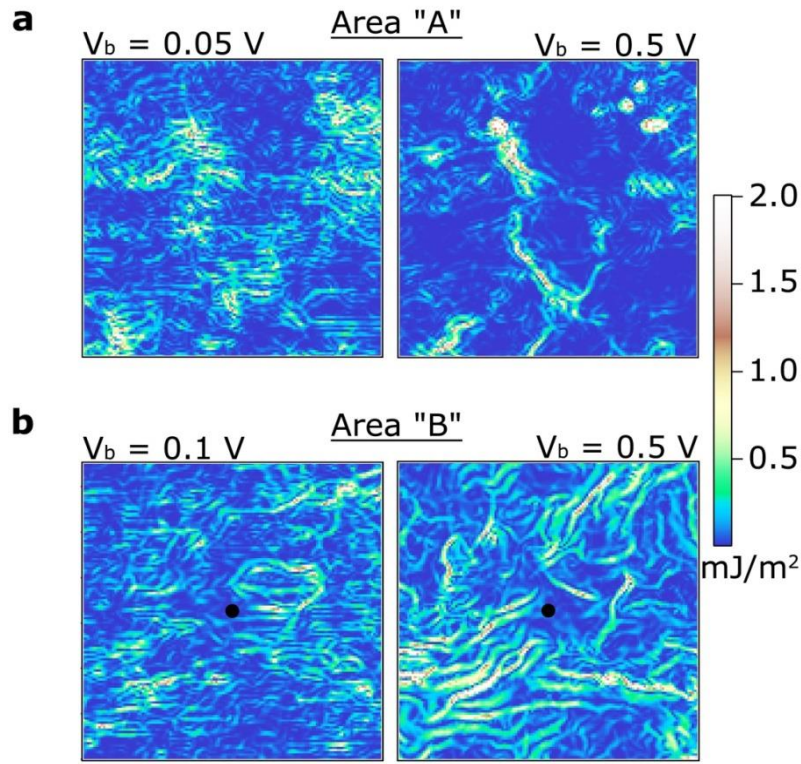

**Fig. 5** Bending energy maps at low and high bias voltages calculated in the white squares drawn in Fig. 1. (a) area A and (b) area B. The black dots are at the same position as the ones in Figure 1b.

## 4 Conclusions

In summary, we have investigated graphene rippling due to the transverse electric field of an STM tip in the proximity of a metallic contact. We showed that the corrugation of graphene depended on the applied bias voltage. Bending energy maps calculated from the topographic images give new insights into the electro-mechanical response for differently stressed graphene areas. The observed rippling enhancement may deeply impact carrier mobility in graphene based nano-electronic devices where transverse electric fields are involved (e.g. suspended top-gate electrodes[38,39]). The experimentally found *H* values can directly be used to estimate the excess resistivity $\Delta\rho \approx n^{1-2H}$ caused by scattering from ripples[9] (*n* is the charge carrier density). This could be used in future devices to tailor the electronic properties of strained graphene membranes[40,41,42] much more efficiently than the usual electrostatic doping.

## Acknowledgements

This work has been performed in the framework of the European project FP7-NMP-2009-SMALL-3 Grenada and the French-Hungarian bilateral programme PHC-Balaton/TÉT_10-1-2011-0752, with the help of the "Plateforme Technologique Amont de Grenoble" and the project DISPOGRAPH financially supported by the "Nanosciences aux limites de la Nanoélectronique" Foundation and the CNRS technology network ReNaTech. The authors acknowledge the grant N° ANR-BLANC-SIMI10-LS-100617-12-01 from the "Agence Nationale de la Recherche" and the cluster micro-nano grant "contact metal-graphene" from the Rhône-Alpes region. We thank M. Morgenstern for insightful discussions. Z.O. acknowledges the post-doctoral fellowships from the CEA and the NEEL Institute in Grenoble as well as the Bolyai János Research Fellowship from the Hungarian Academy of Sciences.

## Notes and references

† Electronic Supplementary Information (ESI) available: STM image of the metal/graphene boundary, as well as Fourier-filtered STM images of graphene to separate the large wavelength roughness from the contribution from rippling. See DOI: 10.1039/C3NR02934D.

1  A. K. Geim and K. S. Novoselov, *Nat. Mater.*, 2007, **6**, 183.
2  K. S. Novoselov, A. K. Geim, S. V. Morozov, D. Jiang, Y. Zhang, S. V. Dubonos, I. V. Grigorieva and A. A. Firsov, *Science*, 2004, **306**, 666.
3  K. S. Novoselov, A. K. Geim, S. V. Morozov, D. Jiang, M. I. Katsnelson, I. V. Grigorieva, S. V. Dubonos and A. A. Firsov, *Nature*, 2005, **438**, 197.
4  K. I. Bolotin, K. J. Sikes, Z. Jiang, M. Klima, G. Fudenberg, J. Hone, P. Kim and H. L. Stormer, *Solid State Commun.*, 2008, **146**, 351.
5  X. Du, I. Skachko, A. Barker and E. Y. Andrei, *Nat. Nanotechnol.*, 2008, **3**, 491.

6   Y.-W. Tan, Y. Zhang, K. Bolotin, Y. Zhao, S. Adam, E. H. Hwang, S. D. Sarma, H. L. Stormer and P. Kim, *Phys. Rev. Lett.*, 2007, **99**, 246803.

7   J.-H. Chen, C. Jang, S. Xiao, M. Ishigami and M. S.Fuhrer, *Nat. Nanotechnol.*, 2008, **3**, 206.

8   J.-H. Chen, C. Jang, S. Adam, M. S. Fuhrer, E. D. Williams and M.Ishigami, *Nat. Phys.*, 2008, **4**, 377.

9   M. I. Katsnelson and A. K. Geim, *Phil. Trans. R. Soc. A*, 2008, **366**, 195.

10  G.-X. Ni, Y. Zheng, S. Bae, H. R. Kim, A. Pachoud, Y. S. Kim, C.-L. Tan, D. Im, J.-H. Ahn, B. H. Hong and B. Özyilmaz, *ACS Nano*, 2012, **6**, 1158.

11  M. Ishigami, J. H. Chen, W. G. Cullen, M. S. Fuhrer and E. D. Williams, *Nano Lett.*, 2007, **7**, 1643.

12  M. L. Teague, A. P. Lai, J. Velasco, C. R. Hughes, A. D. Beyer, M. W. Bockrath, C. N. Lau and N.-C.Yeh, *Nano Lett.*, 2009, **9**, 2542.

13  V. Geringer, M. Liebmann, T. Echtermeyer, S. Runte, M. Schmidt, R. Rückamp, M. C. Lemme and M. Morgenstern, *Phys. Rev. Lett.*, 2009, **102**, 076102.

14  T. Mashoff, M. Pratzer, V. Geringer, T. J. Echtermeyer, M. C. Lemme, M. Liebmann and M. Morgenstern, *Nano Lett.*, 2010, **10**, 461.

15  Z. Wang, L. Philippe and J. Elias, *Phys. Rev. B*, 2010, **81**, 155405.

16  Z. Wang, *Carbon*, 2009, **47**, 3050.

17  J. Sabio, C. Seoánez, S. Fratini, F. Guinea, A. H. Castro Neto and F. Sols, *Phys. Rev. B*, 2008, **77**, 195409.

18  J. C. Meyer, A. K. Geim, M. I. Katsnelson, K. S. Novoselov, T. J. Booth and S. Roth, *Nature*, 2007, **446**, 60.

19  A. Fasolino, J. H. Los and M. I. Katsnelson, *Nat. Mater.*, 2007, **6**, 858.

20  W. Bao, F. Miao, Z. Chen, H. Zhang, W. Jang, C. Dames and C. N. Lau, *Nat. Nanotechnol.*, 2009, **4**, 562.

21  T. A. Witten, *Rev. Mod. Phys.*, 2007, **79**, 643.

22  N. N. Klimov, S. Jung, S. Zhu, T. Li, C. A. Wright, S. D. Solares, D. B. Newell, N. B. Zhitenev and J. A. Stroscio, *Science*, 2012, **336**, 1557.

23  H. Vandepare, M. Pineirua, F. Brau, B. Roman, J. Bico, C. Gay, W. Bao, C. N. Lau, P. M. Reis and P. Damman, *Phys. Rev. Lett.*, 2011, **106**, 224301.

24  V. B. Shenoy, C. D. Reddy, A. Ramasubramaniam and Y. W. Zhang, *Phys. Rev. Lett.*, 2008, **101**, 245501.

25  W. T. Pong and C. Durkan, *J. Phys. D: Appl. Phys.*, 2005, **38**, R329.

26    A. Deshpande, W. Bao, F. Miao, C. N. Lau and B. LeRoy, J. *Phys. Rev. B*, 2009, **79**, 205411.

27    Y. Zhang, V. W. Brar, C. Girit, A. Zettl and M. F. Crommie, *Nat. Phys.*, 2009, **5**, 722.

28    B. J. Hopkins and K. J. Ross, *Brit. J. Appl. Phys.*, 1964, **15**, 89.

29    S. J. Sque, R. Jones and P. R. Briddon, *Phys. Stat. Sol. A*, 2007, **204**, 3078.

30    Y.-J. Yu, Y. Zhao, S. Ryu, L. E. Brus, K. S. Kim and P. Kim, *Nano Lett.*, 2009, **9**, 3430.

31    R. Golizadeh-Mojarad and S. Datta, *Phys. Rev. B*, 2009, **79**, 085410.

32    F. Xia, T. Mueller, R. Golizadeh-Mojarad, M. Freitag, Y.-M. Lin, J. Tsang, V. Perebeinos and P. Avouris, *Nano Lett.*, 2009, **9**, 1039.

33    S. P. Koenig, N. G. Bodetti, M. L. Dunn and J. S. Bunch, *Nat. Nanotechnol.*, 2011, **6**, 543.

34    Z. Zang and T. Li, *J. Appl. Phys.*, 2011, **110**, 083526.

35    Q. Lu, M. Arroyo and R. Huang, *J. Phys. D: Appl. Phys.*, 2009, **42,** 102002.

36    C. Lee, X. Wei, J. W. Kysar and J. Hone, *Science*, 2008, **321**, 385.

37    O. L. Blakeslee, D. G. Proctor, E. J. Seldin, G. B. Spence and T. Weng, *J. Appl. Phys.*, 1970, **41**, 3373.

38    J. Velasco Jr., G. Liu, W. Bao and C. N. Lau, *New J. Phys.*, 2009, **11**, 095008.

39    R. T. Weitz, M. T. Allen, B. E. Feldman, J. Martin and A. Yacoby, *Science*, 2010, **330**, 812.

40    V. M. Pereira and A. H. Castro Neto, *Phys. Rev. Lett.*, 2009, **103**, 046801.

41    V. M. Pereira and A. H. Castro Neto, *Phys. Rev. Lett.*, 2011, **105**, 156603.

42    L. Tapasztó, T. Dumitrică, S. J. Kim, P. Nemes-Incze, C. Hwang and L. P. Biró, *Nat. Phys.*, 2012, **8**, 739.